\documentclass[aps,pre,twocolumn,showpacs,groupedaddress,letterpaper]{revtex4-1}  % for review and submission 
\usepackage{graphicx}  % needed for figures
\usepackage{epstopdf}
\usepackage{subcaption}
\usepackage{ragged2e}

\newcommand{\beq}{\begin{equation}}
\newcommand{\eeq}{\end{equation}}
\newcommand{\bea}{\begin{eqnarray}}
\newcommand{\eea}{\end{eqnarray}}

\begin{document}

\title{Grand-canonical simulation of DNA condensation with two salts, effect of divalent counterion size}

\author{Toan T. Nguyen$^{1,2}$}

\affiliation{$^1$Faculty of Physics, Hanoi University of Science, Vietnam National University, 334 Nguyen Trai Street, 
Thanh Xuan, Hanoi, Vietnam\\
$^2$School of Physics, Georgia Institute of Technology, 837 State Street, Atlanta, Georgia 30332-0430, USA\\
}

\date{\today}

\begin{abstract}
The problem of DNA$-$DNA interaction mediated by divalent counterions 
is studied using a generalized Grand-canonical
Monte-Carlo simulation for a system of two salts.
The effect of the divalent counterion
size on the condensation behavior of the DNA bundle is investigated. Experimentally,
it is known that multivalent counterions have strong effect on the DNA
condensation phenomenon. While tri- and tetra-valent counterions are shown to 
easily condense free DNA molecules in solution into toroidal bundles, 
the situation with divalent
counterions are not as clear cut. Some divalent counterions like Mg$^{+2}$ are
not able to condense free DNA molecules in solution, while some like Mn$^{+2}$
can condense them into disorder bundles. In restricted environment such as in
two dimensional system or inside viral capsid, Mg$^{+2}$ can have strong
effect and able to condense them, but the condensation varies qualitatively
with different system, different coions. It has been suggested that
divalent counterions can induce attraction between DNA molecules but
the strength of the attraction is not strong enough to condense free DNA
in solution. However, if the configuration entropy of DNA is restricted,
these attractions are enough to cause appreciable effects. The variations
among different divalent salts might be due to the hydration effect of the divalent counterions.
In this paper, we try to understand this variation using a very simple
parameter, the size of the divalent counterions. We investigate how
divalent counterions with different sizes can leads to varying qualitative
behavior of DNA condensation in restricted environments. 
Additionally a Grand canonical Monte-Carlo method for simulation of
systems with two different salts is presented in detail.
\end{abstract}

% insert suggested PACS numbers in braces on next line
% 82.39.Pj Nucleic acids, DNA and RNA bases 
% 87.14.gn RNA 
% 87.15.N- Properties of solutions of macromolecules 
% 87.19.rm Structure 
% 81.16.Dn Self-assembly
% 87.16.A- Theory, modeling, and simulations 
% insert suggested keywords - APS authors don't need to do this
%\keywords{}

%\maketitle must follow title, authors, abstract, \pacs, and \keywords
\maketitle

\section{Introduction}

The problem of DNA condensation in the presence of multivalent counterions
has seen a strong revival of interest in recent years. This is 
because of the need to develop effective ways of
gene delivery for the rapidly growing field of genetic therapy. 
DNA viruses such as bacteriophages
provide excellent study candidates for this purpose. 
One can package genomic DNA into viruses, 
then deliver and release the molecule into targeted individual cells. 
Recently there is a large biophysics literature dedicated to the problem of 
DNA condensation (packaging and ejection) inside bacteriophages 
(for a review, see Ref. \onlinecite{GelbartVirusReview2009}).

% It is well-known that the persistence
% length $l_p$ of DNA is about 50 nm, comparable to or even larger than the inner 
% diameter of the viral capsid. 
% The genome of a typical bacteriophage is about 10 microns
% or 200 persistence lengths. Thus, the DNA molecule is 
% considerable bent and strongly confined inside the viral capsid,
% resulting in a substantially pressurized capsid with 
% internal pressure as high as 
% 50 atm \cite{Bustamante01,Gelbart03,Gelbart2003,Harvey07}.
% It has been suggested that this
% pressure is the main driving force for the ejection of the
% viral genome
% into the host cell when the capsid tail binds to
% the receptor in the cell membrane and subsequently opens the capsid.
% This idea is supported by various experiments both
% \textit{in vivo} and \textit{in vitro}
% \cite{Santamaria04,Gelbart03,Black89,Murialdo91,Gelbart2003,Phillips05, Gelbart04,Knobler08}.
% The \textit{in vitro} experiments additionally revealed possibilities of
% controlling
% the ejection of DNA from bacteriophages. One example is the addition of
% PEG (polyethyleneglycol), a large molecule that is incapable of penetrating
% the viral capsid. A finite PEG concentration in solution
% produces an apparent osmotic pressure on the capsid.
% This in turn leads to a reduction
% or even complete inhibition of the ejection of DNA.

Because DNA is a strongly charged molecule in aqueous solution, electrostatics
and the screening condition of the solution
play an important role in the structure and functions of DNA systems.
Specifically, the condensation of DNA molecules is strongly
influenced by the counterion valence \cite{Parsegian92,Hud01,HoangTorroidJCP2014,GrasonPRL2010}. 
While tri- and tetra-valent counterions are shown to 
easily condense free DNA molecules in solution into toroidal bundles, 
the situation with divalent
counterions are not as clear cut. 
Some divalent counterions like Mg$^{+2}$ are
not able to condense free DNA molecules in solution, while some like Mn$^{+2}$
can condense them into disorder bundles.  
Similarly, strong electrostatic effect is also observed for DNA condensation in a restricted
environment such as inside a viral capsid. By varying the salinity of solution,
 one can vary the amount of DNA ejected from viruses. Interestingly, monovalent
counterions such as Na$^{+1}$ have negligible effect on 
the DNA ejection process \cite{Gelbart03}.
In contrast, multivalent counterions ($Z-$ions for short)
such as Mg$^{+2}$, CoHex$^{+3}$,
Spd$^{+3}$ or Spm$^{+4}$ exert
strong and non-monotonic effects \cite{Knobler08}. There is an optimal counterion
 concentration, $c_{Z,0}$, where the least DNA genome is ejected from the phages. 
For counterion concentration, $c_Z$, higher or lower than this optimal concentration,
more DNA is ejected from phages. The case of divalent counterions is more marginal.
 The non-monotonicity is observed for MgSO$_4$ salt but not for MgCl$_2$ salt up to 
the concentration of 100mM. Such ion specificity
for the case of divalent salts also present in condensation of DNA in free solution. 
\cite{Parsegian92}. 

The non-monotonic influence of multivalent counterions on DNA ejection from viruses 
is expected to have the same physical origin as the phenomenon of
reentrant DNA condensation in free solution in the presence
of counterions of tri-, tetra- and higher valence
\cite{NguyenJCP2000,SaminathanBiochem1999,LivolantBJ1996,NetzLikeChargedRods,GelbartPhysToday}.
Although, divalent counterions are known to condense
DNA only partially in free solution \cite{Parsegian92,Hud01}, DNA virus
provides a unique experimental setup. The constraint of
the viral capsid strongly eliminates configurational entropic cost of
packaging DNA. This allows divalent counterions to influence
 DNA condensation similar to that of trivalent/tetravalent
counterions. Indeed, DNA condensation by divalent counterions has
also been observed in another environment where DNA configuration is 
constrained, namely the 
condensation of DNA in two dimensional systems \cite{Koltover2000}. 
For virus systems, theoretical fitting suggests that the DNA is neutralized 
at $c_{Z,0}\approx 75$mM for divalent counterions, and the short$-$range DNA 
attraction at this concentration is $-0.004k_BT$ per nucleotide base \cite{NguyenJCP2011,NguyenJBP2013}. 

In this paper, we study the problem of DNA condensation in the
presence of divalent counterions using computer simulations.
The simulation method developed by our groups in
Ref. \onlinecite{NguyenPRL2010,NguyenJBP2013} is used, expanded and the influence of
the ion size on the strength of DNA$-$
DNA interaction mediated by divalent counterions is investigated
\cite{DNAreentrantfootnote}.
The Grand Canonical Monte Carlo simulation for a system of two salts
is presented in detail.
The electrostatic contribution to the free energy of packaging DNA into
 bundles is calculated from simulation. It is shown that, if only the non-specific electrostatic contribution 
is included, divalent counterions can indeed induce DNA reentrant 
condensation like those observed for higher counterion valences.
However, correlations among divalent counterions are not strong enough
to de-condense DNA bundles.
As already mentioned, 
experimental results
also show that there is a ion specific effect. 
As a first step taken to study this ion specific
effect, the DNA$-$DNA effective interaction is calculated from
simulation for three different counterion sizes. It is shown that 
varying counterion sizes can have significant impact on DNA
condensation pictures, which can explained some variations among DNA condensation
experiments with Mg$2+$, or Mn$2+$ counterions.

The paper is organized as follows. In Sec. \ref{sec:GCMC}, the Grand-canonical Monte-Carlo is 
formulated to simulate a system of two salts (a divalent salts and a fixed monovalent salt from buffer solution).
In Sec. \ref{sec:model}, the model of our system
and various physical parameters used in the simulation are presented in details.
% In Sec.
% \ref{sec:GCMC} we briefly explain the method of Expanded Ensemble
% Grand Canonical Monte$-$Carlo simulation 
% for calculating the free energy of DNA bundle. This method has been 
% modified in this paper to simulate system with two different
% salts present. 
In Sec. \ref{sec:discussion}, the
results are presented and  their relevance to available experimental data is discussed.
We conclude in Sec. \ref{sec:conclusion}.

\section{Grand canonical Monte$-$Carlo Simulation
for mixture of two salts\label{sec:GCMC}}

In practical situation, the DNA bundle is in equilibrium
with a water solution containing free mobile ions at 
given concentrations. Therefore we simulate the system using 
Grand Canonical Monte-Carlo (GCMC) simulation.
The number of ions is not constant during the simulation. Instead
their chemical potentials are fixed. These chemical potentials
are chosen in advance by simulating a DNA$-$free salt solution
and adjusting them so that the solution has the correct
ion concentrations. 
Another factor that complicates the
simulation of DNA condensation phenomenon arises from
the fact that there are both monovalent
and divalent salts in solution in experiments. 
At very low concentration of divalent counterions, $c_Z$,
DNA is screened mostly by monovalent counterions. 
To properly simulate the DNA bundle at this low $c_Z$ limit, 
and to properly capture the screening of electrostatic
interactions among divalent counterions by monovalent ones,
both salts are included in the simulations .

To simulate two different salts present in our system,
the standard GCMC method for ionic solution [\onlinecite{CohenGCMC}]
is generalized to simulate of a system containing a mixture of 
both multivalent and monovalent salts.
For simplicity, we assume both salts have the same coion (for example,
Cl$^{-}$).
Thus, a state $i$ of the system is characterized by
the locations of $N_{iZ}$ multivalent counterions, 
$N_{i+}$ monovalent counterions and $N_{i-}$ coions. 
In the grand canonical ensemble of unlabeled particles, 
the probability of such state is given by
\beq
\pi_i = \frac{1}{\cal Z} 
  \frac{1}{\Lambda_Z^{3N_{iZ}} \Lambda_+^{3N_{i+}} \Lambda_-^{3N_{i-}}}
\exp \left[\beta(\mu_ZN_{iZ}+\mu_+N_{i+}+\mu_-N_{i-})-\beta U_i\right]
\label{GCMCpii}
\eeq
Here, $\cal Z$ is the grand canonical partition function,
$\beta=1/k_BT$, $\Lambda_{Z,+,-}\equiv h/\sqrt{2\pi m_{Z,+,-} k_BT}$,
$U_i$ is the interaction energy of the state $i$,
and $\mu_{Z,+,-}$ are the chemical potentials of the
multivalent counterions, of the monovalent counterions and
of the coions respectively.

In a Monte Carlo simulation, a Markov chain of system states $i$
is generated with a limiting probability distribution proportional
to $\pi_i$. This chain is defined by a probability
$p_{ij}$ of transitions from state $i$ to state $j$. A sufficient
condition for the Markov chain to have the correct limiting
distribution is:
\beq
\frac{p_{ij}}{p_{ji}} = \frac{\pi_j}{\pi_i}
\eeq
As usual, at each step of the chain, a ``trial" move to change
the system from state $i$ to state $j$ is attempted with probability $q_{ij}$
and is accepted with probability $f_{ij}$. Clearly,
\beq
p_{ij} = q_{ij} f_{ij}
\eeq
It is convenient to regard the simulation box as consisting of
$V$ discrete sites ($V$ is very large). Then for a trial move
where $\nu_{\alpha}$ particles of species $\alpha$ are added
to the system:
\beq
q_{ij} = \frac{1} {V^{\nu_\alpha} \nu_\alpha!}
\eeq
Conversely, if $\nu_{\alpha}$ particles of species $\alpha$ are
removed from the system:
\beq
q_{ij} = \frac{ (N_\alpha - \nu_\alpha)! } {N_\alpha! \nu_\alpha!}
\label{GCMCqijdelete}
\eeq

Putting everything together, equations 
(\ref{GCMCpii})$-$(\ref{GCMCqijdelete}) give
us a recipe to calculate the Metropolis acceptance
probability of a particle insertion/deletion move in
GCMC simulation. For example, if in a transition
from state $i$ to state $j$, a multivalent salt molecule
(one $Z-$ion and $Z$ coions) is added to the system,
the Metropolis probability of acceptance of such
move can be chosen as:
\beq
f_M = \min\{1, ~ f_{ij}/f_{ji} \}
\eeq
where
\beq
\frac{f_{ij}}{f_{ji}} = \frac{B_Z}
  {(N_{iZ}+1)(N_{i-}+1)...(N_{i-}+Z)}
   \exp[ \beta (U_i-U_j) ],
\eeq
with
\beq
\label{eq:BZ}
B_Z = \exp(\beta\mu_{Z,\mbox{salt}}) \frac{V^{Z+1}}{\Lambda_Z^3 \Lambda_-^{3Z}},
\eeq
and 
\beq
\mu_{Z,\mbox{salt}} = \mu_Z + Z\mu_-
\eeq
is the combined chemical potential of a multivalent salt molecule.

On the other hand, if a multivalent salt molecule 
(one $Z-$ion and $Z$ coions) is removed from the system,
\beq
\frac{f_{ij}}{f_{ji}} = \frac{N_{iZ}N_{i-}...(N_{i-}-Z+1)}
  {B_Z} \exp[ \beta (U_i-U_j) ],
\eeq
Similarly, for addition a monovalent salt molecule (one
monovalent counterion and one coion) in transition
from state $i$ to state $j$,
\beq
\frac{f_{ij}}{f_{ji}} = \frac{B_1}
  {(N_{iZ}+1)(N_{i-}+1)}
   \exp[ \beta (U_i-U_j) ],
\eeq
with
\beq
\label{eq:B1}
B_1 = \exp(\beta\mu_{1,\mbox{salt}}) \frac{V^2}{\Lambda_+^3 \Lambda_-^3},
\eeq
and 
\beq
\mu_{1,\mbox{salt}} = \mu_+ + \mu_-
\eeq
is the combined chemical potential of a monovalent salt molecule.
For a ``trial" move where a monovalent salt molecule 
is removed from the system,
\beq
\frac{f_{ij}}{f_{ji}} = \frac{N_{i+}N_{i-}}{B_1}
   \exp[ \beta (U_i-U_j) ],
\eeq

Because we are trying to simulate a mixture of salts,
to improve the system relaxation and to
improve the sampling of the system's phase space,
one can also make a ``trial" move where one $Z-$ion is
added to the system and $Z$ monovalent counterions
are removed the system. For such move, it is easy to show that
\beq
\frac{f_{ij}}{f_{ji}} = \frac{B_1^Z N_{i+}...(N_{i+}-Z+1)}
  {B_Z(N_{iZ}+1)}
   \exp[ \beta (U_i-U_j) ],
\eeq
Vice versa, for a ``trial" move where one $Z-$ion is
removed from the system and $Z$ monovalent counterions
are added to the system,
\beq
\frac{f_{ij}}{f_{ji}} = \frac{B_Z N_{iZ}}
  {B_1^Z (N_{i+}+1)...(N_{i+}+Z)}
   \exp[ \beta (U_i-U_j) ].
\eeq

Note that because the system maintains charge neutrality
in all particle addition/deletion moves, instead
of using 3 different chemical potentials,
$\mu_{Z,+,-}$, to simulate
the system, only two combined chemical potentials, 
$\mu_{Z,\mbox{salt}}$ and $\mu_{1,\mbox{salt}}$, are
actually needed. In our actual implementation,
the dimensionless parameters $B_Z$ and $B_1$,
Eqs. (\ref{eq:B1}) and (\ref{eq:BZ}), are used 
instead of the chemical potentials themselves
to simulate the DNA system. The values of these
parameters for different mixtures of 
divalent and monovalent salts are listed in Sec. 
\ref{sec:simulationDetail}, Table I.

Lastly, beside particle addition/deletion moves, one also
try standard particle translation moves. They are carried
out exactly like in the case of a canonical Monte-Carlo simulation.
In a ``trial" move from state $i$ to state $j$,
an ion is chosen at random and is moved to a random
position in a volume element surrounding its original
position. The standard Metropolis probability
is used for the acceptance of such ``trial" move:
\beq
f_M = \min \{1, ~ \exp[\beta(U_i-U_j)]\}.
\eeq

\section{The simulation model\label{sec:model}}
\label{sec:simulationDetail}
\begin{figure}[ht]
\resizebox{8cm}{!}{\includegraphics{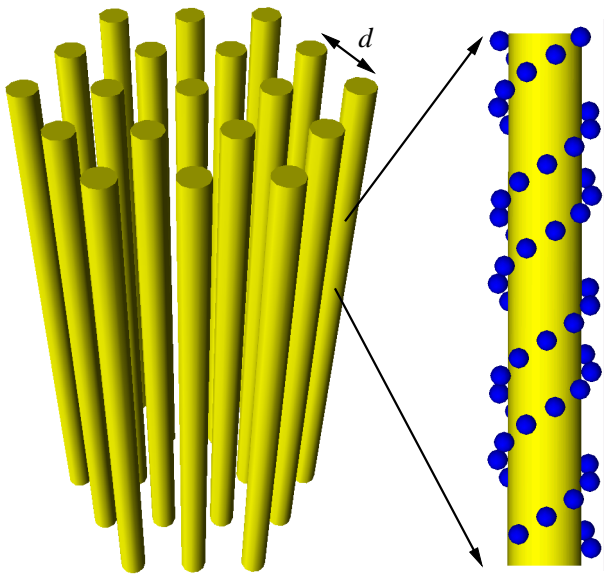}}
\caption{(Color online) A DNA bundle is modeled as a hexagonal lattice
with lattice constant $d$. Individual
DNA molecule is modeled as a hard-core cylinder with negative charges
glued on it according to the positions of nucleotides of a B$-$DNA structure.
}
\label{fig:DNA}
\end{figure}

We model the DNA bundle in hexagonal packing as a number of DNA molecules 
arranged in parallel along the $Z$-axis. In the horizontal plane, the DNA 
molecules form a two dimensional hexagonal lattice with lattice constant
$d$ (the DNA$-$DNA interaxial distance) (Fig. \ref{fig:DNA}). 
Individual DNA molecule
is modeled as an impenetrable cylinder with negative charges glued on it. 
The charges are positioned in accordance with the locations of nucleotide groups
 along the 
double-helix structure of a B$-$DNA. The hardcore cylinder
has radius of 7\AA. The negative charges are hard spheres
of radius 2\AA, charge $-e$ and lie at a 
distance of 9\AA\  from the DNA axis. This
gives an averaged DNA radius, $r_{DNA}$, of 1nm. The solvent water is treated as
a dielectric medium with dielectric constant $\varepsilon = 78$
and temperature $T=300^oK$. The positions of DNA molecules are fixed in space. 
This mimics the constraint on DNA configurational entropy inside viruses and 
other experiments of DNA condensation using divalent counterions in restricted
environment. The mobile 
ions in solution are modeled as hard spheres with unscreened Coulomb interaction 
(the primitive ion model).
The coions have radius of $\sigma_-=2$\AA\ and charge $-e$. 
The divalent counterions have radius of $\sigma_Z =$ 2.0, 2.5, or 3.0\AA\ 
and charge $+2e$. The interaction between two ions
$\alpha$ and $\beta$ with radii $\sigma_{\alpha,\beta}$ and charges $Q_{\alpha,\beta}$ is given by
\beq
U = \left\{
  \begin{array}{l l}
    Q_\alpha Q_\beta/\varepsilon r_{\alpha\beta}& \quad 
\mbox{if $r_{\alpha\beta} > \sigma_\alpha + \sigma_\beta$}\\
    \infty & \quad \mbox{if $r_{\alpha\beta} < \sigma_\alpha + \sigma_\beta$}\\ 
  \end{array}
    \right.
\eeq
where $r_{\alpha\beta}=|\mathbf{r}_\alpha-\mathbf{r}_\beta|$ is the distance
between the ions. 

The simulation is carried out using the periodic boundary condition. Unless
explicitly stated, a periodic simulation cell with 
$N_{DNA} = 12$ DNA molecules in the horizontal $(x,y)$ plane and 3 full helix 
periods in the $z$ direction is used. The dimensions of the box are $L_x = 3d$,
 $L_y=2\sqrt{3}d$ and $L_z = 102$\AA. This gives, for the volume of the simulation
box,
\beq
V_{\mbox{cell}} = 612 \sqrt{3}\ d^2\ \mbox{\AA}^3
\eeq
The long-range electrostatic interactions
between charges in neighboring cells are treated using the
Ewald summation method. %\cite{EwaldSum}. 
In Ref. \cite{Nordenskiold95,NordenskioldJCP86}, it is shown that
the macroscopic limit is reached when $N_{DNA} \geq 7$. Our simulation cell
contains 12 DNA helices,
hence it has enough DNA molecules to eliminate the finite size effect.
Test runs with 1, 4, 7 and 12 DNA molecules are carried out to verify that
this is indeed the case. 
%These smaller simulations are also used
%to check the correctness of our computer program by reproducing the results
%of DNA systems studied in Ref. \cite{Nordenskiold95,NordenskioldJCP86}
%in appropriate limits.

% In this paper, we simulate DNA bundles at varying 
% counterion concentration $c_Z$. 
% The limit of small $c_Z$
% poses a non-trivial challenge. If we simulate DNA bundle in the presence of
% only divalent salt, then even at $c_Z=0$, 
% there would be non-zero amount of divalent counterions inside the bundle
% due to the charge neutrality requirement. This situation is
% clearly non-physical. In reality, there always a finite amount of monovalent
% counterions from the buffer solution or from the deprotonization
% of DNA bases. According to the mass action law,
% when $c_Z$ is smaller than a certain value, the monovalent counterions
% will replace the divalent ones in neutralizing DNA charges.
% Therefore, in this paper, to properly simulate
% the DNA bundle at small $c_Z$ limit, 
% we include {\em both} divalent and monovalent salts in the simulation. 
% The standard GCMC simulation method is generalized to this system by allowing
%  insertion/removal of both kinds of salts in a simulation run. For simplicity,
% they both assumed to have the same coions. In addition to the chemical potential 
% of a divalent salt molecule, $\mu_{+Z~salt}$, we also use the chemical potential
%  of a monovalent salt molecule,
% %
% $
% \mu_{+1~ salt} = \mu_{+1} + \mu_{-1}
% $,
% %
% in the Metropolis criteria.

As mentioned above, the DNA bundle is simulated in equilibrium with a bulk solution
containing two salt concentrations: a varying bulk multivalent 
counterion concentrations $c_Z$ and a fixed bulk concentration
of monovalent salt, $c_1 = 50$mM. The detail
implementation of the GCMC method for this case is described in section II.
%The number of ions is not constant during the simulation. Instead
%their chemical potentials are fixed. These chemical potentials
%are chosen in advance by simulating a DNA$-$free salt solution
%and adjusting them so that the solution has the correct
%ion concentrations. 
% To simulate two different salts present in our system,
% the standard GCMC method for ionic solution [\onlinecite{CohenGCMC}]
% is generalized to simulate of a system containing a mixture of 
% both multivalent and monovalent salts \cite{NguyenJBP2013}. 
%Following the notation of 
%Ref. \onlinecite{CohenGCMC}, 
In simulation, the chemical potential of each salt is set by 
fixing the parameters $B_{1,Z}$ given by Eq. (\ref{eq:BZ}, \ref{eq:B1}).
% %
% \beq
% \label{eq:BZ}
% B_Z = \exp(\beta\mu_{Z,\mbox{salt}}) \frac{V^{Z+1}}{\Lambda_Z^3 \Lambda_-^{3Z}},
% \eeq
% %
In Table \ref{table:mu}, various values for the parameters 
$B_Z^*$ and $B_1^*$ that are used in this work for divalent counterion size of 2\AA\ are shown.
These values are listed for a reference volume $V_{\mbox{cell}}^*$ that  
is chosen to have the same dimensions as that of a DNA bundle system 
with $d=50$\AA, so $V_{\mbox{cell}}^* \simeq 2.65\times 10^6$ \AA$^3$. 
For a simulation system where $d$ is different from 50\AA, 
the parameters $B_Z$ and $B_1$ are scaled correspondingly:
\beq
B_Z (d) = B_Z^*\left(\frac{d}{50\mbox{\AA}}\right)^{2Z+2},
B_1 (d) = B_1^*\left(\frac{d}{50\mbox{\AA}}\right)^4. \nonumber
\eeq
\begin{center}
\begin{table}
\begin{tabular}{c|c||c|c|c}
\hline
$B_Z^*$ & $B_1^*$ & $c_Z$ (mM) & $c_1$ (mM) & $P_b$ (atm)\\
\hline
$0.744\times 10^5$ & $0.612\times 10^4$ & $13.9\pm 3.0$ & $50.0\pm 5.6$ & $3.183\pm 0.001$\\
$2.568\times 10^5$ & $0.808\times 10^4$ & $29.9\pm 3.4$ & $50.2\pm 4.9$ & $4.17\pm 0.01$\\
$14.48\times 10^5$ & $1.306\times 10^4$ & $74.6\pm 6.2$ & $50.1\pm 5.3$ & $6.874\pm 0.006$\\
$26.43\times 10^5$ & $1.580\times 10^4$ & $99.8\pm 5.7$ & $50.3\pm 5.4$ & $8.391\pm 0.006$\\
$56.67\times 10^5$ & $2.128\times 10^4$ & $150.2\pm 8.4$ & $50.6\pm 6.7$ & $11.42\pm 0.02$\\
$323.82\times 10^5$ & $3.715\times 10^4$ & $299.6\pm 11.2$ & $49.4\pm 6.8$ &$20.81\pm 0.04$\\
$1302.73\times 10^5$ & $6.601\times 10^4$ & $507.1\pm 13.6$ & $50.3\pm 6.9$ & $35.0\pm 0.1$\\
\hline
\end{tabular}
\caption{The parameters, $B_Z^*$ and $B_1^*$, of the salts used in the simulation
for the reference volume $V_{\mbox{cell}}^*\simeq 2.65\times 10^6$ \AA$^3$ (see
text for detail). Columns 3 and 4 show the corresponding salt 
concentrations of the simulated DNA$-$free bulk solution. 
Column 5 shows the total pressure of the bulk solutions
obtained from simulation.}
\label{table:mu}
\end{table}
\end{center}
In columns 3 and 4 of table I, the resultant salt concentrations, $c_Z$ and
$c_1$, of the DNA$-$free solution obtained from our GCMC simulations
are listed. The divalent salt concentration is varied
from 14 mM to 507 mM while the monovalent salt concentration is
kept at approximately 50 mM. Typical standard deviations in the
concentration is about 10\% in our simulation. This relative error is
in line with previous GCMC simulations of primitive electrolytes
\cite{CohenGCMC}. Note that, even though $c_1$ is kept constant, 
$B_1^*$, (and correspondingly the monovalent salt
chemical potential $\mu_{1,\mbox{salt}}$,) is not a constant but 
actually increases with $c_Z$. This is expected because higher
$c_Z$ leads to higher free energy cost of adding a monovalent
salt to the system.

For each simulation run, about 500-1000 million MC moves are 
carried out depending on the average number of ions in the
system. To ensure thermalization, about 50 million initial moves 
are discarded before doing statistical analysis of 
the result of the simulation.

In this paper, we are concerned with calculating the
``effective" DNA$-$DNA interaction, and correspondingly
the free energy of assembling DNA bundle. In general, this
is not a trivial task for a Monte-Carlo simulation
because the entropy cannot be calculated explicitly. 
To overcome this problem, the Expanded Ensemble
method \cite{Nordenskiold95} is implemented. This method allows us to
calculate the difference of the system free energies at
different volumes by sampling these volumes simultaneously 
in a simulation run. By sampling two nearly equal volumes, 
$V$ and $V+\Delta V$, and calculate the free energy difference 
$\Delta \Omega$, we can calculate the total pressure of the system:
\beq
P(T,V,\{\mu_\nu\}) = - \left.
      \frac{\partial \Omega(T,V,\{\mu_\nu\})}{\partial V}
    \right|_{T,\{\mu_\nu\}} 
    \simeq -\frac{\Delta \Omega}{\Delta V}
\label{eq:pressure}
\eeq
Here $\{\mu_\nu\}=\{\mu_Z, ~\mu_1,~\mu_{-1}\}$ are the set of chemical potentials of 
different ion species.  The osmotic pressure of the DNA bundle is 
then obtained by subtracting the total pressure of the 
bulk DNA$-$free solution, $P_b(T,V,\{\mu_\nu\})$, 
from the total pressure of the DNA system:
\beq
P_{osm} (T,V, \{\mu_\nu\}) = P(T,V, \{\mu_\nu\}) - P_b(T,V,\{\mu_\nu\})
\nonumber
\eeq
The total pressure of the bulk solution, $P_b(T,V,\{\mu_\nu\})$, 
needs to be calculated only once for each set of salt concentrations,
$c_Z$ and $c_1$. 
For reference purpose, their values are listed
in column 5 of Table \ref{table:mu}.

All simulations are done using the physics simulation library
SimEngine develop by one of the author (TTN).
This library use OpenCL and OpenMP extensions of the C programming
language to distribute computational workloads on multi-core
CPU and GPGPU to speed up the simulation time. Both molecular dynamics
and Monte-Carlo simulation methods are supported. In this paper
the Monte-Carlo module of the library is used.

\section{Result and Discussion\label{sec:discussion}}

\subsection{Counterion mediated DNA$-$DNA interactions and
 the DNA packaging free energy}

\begin{figure}[ht]
\resizebox{8cm}{!}{\includegraphics{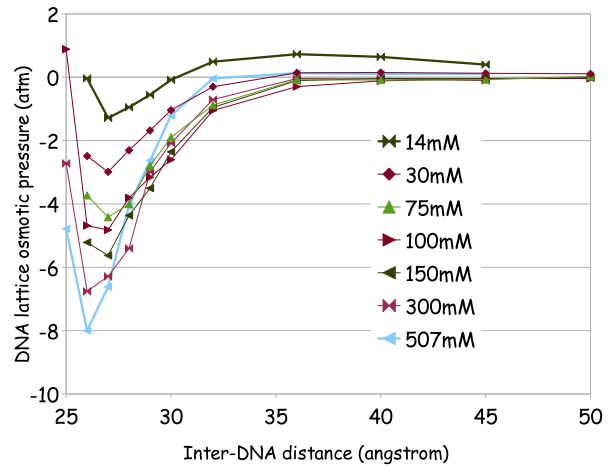}}
\caption{(Color online) The osmotic pressure of the DNA bundle
as function of the interaxial DNA distance $d$ for different
divalent counterion concentration $c_Z$ shown in the inset. 
The solid lines are guides to the eye. The counterion radius is
2.0 \AA
}
\label{fig:posm20}
\end{figure}

In Fig. \ref{fig:posm20},
the osmotic pressure of DNA bundle
at different $c_Z$ is plotted as a 
function of the interaxial DNA distance, $d$ for the case the counterion size is 2\AA. 
Because this osmotic pressure is directly related to
the ``effective'' force between DNA molecules at that interaxial
distance \cite{Nordenskiold95,NordenskioldJCP86}, 
this figure also serves as a plot of DNA$-$DNA interaction. 
As one can see, when $c_Z$ is greater than a value 
around 20mM, there is a short$-$range attraction between two DNA molecules as 
they approach each other. This is the well-known phenomenon of like-charge attraction
between macroions \cite{NetzLikeChargedRods,GelbartPhysToday,NguyenRMP2002}. 
It is the result of the electrostatic
correlations between counterions condensed on the surface of
each DNA molecule. The attraction appears when the distance between these
surfaces is of the order of the lateral separation
between counterions (about 14\AA\ for divalent counterions). The maximal attraction 
occurs at the distance $d\simeq 27$\AA, in good agreement with various theoretical and 
experimental results \cite{Parsegian92,Phillips05}. 
For smaller $d$, the DNA-DNA interaction
experiences sharp increase. This can be understood as the result
of the hardcore repulsion
between the counterions. 

\begin{figure}[ht]
\resizebox{8cm}{!}{\includegraphics{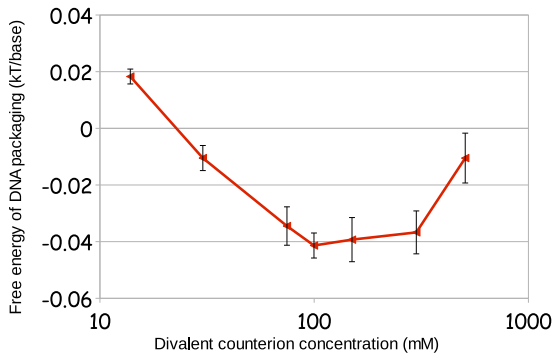}}
\caption{(Color online) The free energy of packaging DNA molecules
into hexagonal bundles as a
function of the divalent counterion concentrations.
The points are results of numerical integration of $P_{osm}$ from
Fig. \ref{fig:FvC}. 
}
\label{fig:FvC}
\end{figure}

From the P-V curve, we can also can calculate the free energy, 
$\mu_{\mbox{DNA}}$, of packaging DNA into bundles. This free 
energy is nothing but the difference between the free energy of 
a DNA molecule in a bundle and that of an  individual DNA molecule
in the bulk solution ($d=\infty$). It can be calculated by 
integrating the pressure with the volume of the bundle.
Per DNA nucleotide base, the packaging free energy is given by:
\bea
\mu_{\mbox{DNA}}(d) &=& \frac{l}{L_z N_{\mbox{DNA}}}
  \int_\infty^d P_{osm}(d') dV 
  \nonumber \\
&=& \frac{l}{N_{\mbox{DNA}}}
  \int_\infty^d P_{osm}(d') \frac{2 L_x L_y}{d'} dd' 
\eea
here $l=1.7$\AA\  is the distance between DNA nucleotides along the
axis of the DNA. The numerical result for $\mu_{\mbox{DNA}}(d^*)$ at
the optimal bundle lattice constant $d^*$ is plotted in Fig. 
\ref{fig:FvC} as function of the $c_Z$. Due to the limitation of computer simulations, 
the numerical integration is performed up to the distance $d=50$\AA\ only. 
However, this will not change the conclusion of this paper because the omitted 
integration from $d=50$\AA\ to $d=\infty$ only 
gives an almost constant shift to $\mu_{\mbox{DNA}}$.
As evident from Fig. \ref{fig:FvC},
the non-monotonic dependence of the electrostatic contribution to DNA 
packaging free energy is clearly shown.
There is an optimal concentration, $c_{Z,0}$, where the
free energy cost of packaging DNA is lowest. 
It is negative indicating the tendency of the divalent
counterions to condense the DNA. At smaller or larger concentrations
of the counterions, the free energy cost of DNA packaging
is higher. These results are consistent with 
the correlation theory of DNA reentrant condensation 
by multivalent counterions \cite{Shklovskii1999,NguyenRMP2002,NguyenJCP2000}
and the experiment results on ejecting DNA from bacteriophage under varying
counterion concentrations \cite{Knobler08}.
% For small $c_Z$, DNA molecules are 
% undercharged ($\eta^* < 0$). For large $c_Z$, 
% DNA molecules are overcharged ($\eta^* >0$). To condense the DNA molecules,
%  one has to overcome the Coulomb repulsion between them. Therefore, the 
% free energy cost of packaging is positive. For $c_Z\approx c_{Z,0}$, 
% the DNA molecules are almost neutral, 
% $\eta^*\approx 0$. The Coulomb repulsion is negligible and the free energy cost
%  of condensing DNA molecules is lowest. Furthermore, the like-charge attraction
%  among DNA molecules mediated by the counterions \cite{NetzLikeChargedRods} 
% is dominant in this concentration range, causing the electrostatic packaging
%  free energy to become negative. 
However, it must be stated, unlike the condensation with counterions of higher
valence \cite{Parsegian92,NguyenJCP2000,lemayNature2007}, 
the divalent counterions in our simulation
are not able to decondense the DNA bundle within the range of concentration
considered. The free energy doesnot become positive beyond $c_{Z,0}$.
This is in line with experimental results \cite{Parsegian92}.
 
 Figure \ref{fig:FvC} gives 
the short$-$range attraction among DNA  molecules to be $-0.04k_BT$/base. 
This is  larger than the fitted value obtained from 
the viral DNA ejection experiments \cite{NguyenJCP2011}. 
There are many factors that lead to this quantitative discrepancy.
Our main approximation is that in the simulation, the position of 
the DNA cylinders are straight with
infinite bending rigidity. Inside viruses, DNA are bent, and the
configuration entropy of the DNA are not necessary zero, and there is
not a perfect hexagonal arrangement of DNA cylinder with fixed inter$-$DNA
distance. We also neglect the contribution from the region $d > 50$\AA\
in our integration. The physical parameters of 
the system such as ion sizes, DNA orientations (twisting, frustrations),... 
\cite{BruinsmaPRLRod,NordenskioldPRL98,GrasonPRL2010} can also affect the strength
of DNA$-$DNA short range attraction.
All these factors are expected to reduce the attraction between
the DNA compared to our idealized simulation. 
Nevertheless, the non-monotonic 
electrostatic influence of divalent counterions on DNA-DNA ``effective" 
interaction is clearly demonstrated in our idealized simulation.

% Another important to note is that, for simplicity, we simulate
% the system with monovalent coions. The neutralizing concentration
% $c_{Z,0}$ is about 100mM from our simulation in this case. 
% In the experiment setup, the MgSO$_4$ salt (divalent coions) 
% shows a minimum in the amount of DNA ejection from viral capsid
% at about 64mM. The data for MgCl$_2$ salt (monovalent coions) seems 
% to suggest a minimum in DNA ejection at about 100mM, about
% the same as our simulation. It would be interesting to see what 
% happens at higher concentrations in the later case.

\subsection{Role of finite size of counterions}
\label{sec:ionSize}

In all the systems simulated so far, the radius of the divalent
counterion is fixed at 2.0\AA. The results agree qualitatively and 
semi-quantitatively with some of the experimental results of DNA ejection from capsid
with MgSO$_4$ salt. However, experimental results
also show that there is an ion specific effect. There
are some significant differences in condensations of free DNA, 
condensations of DNA inside viruses when different divalent salts
such as MgSO$_4$, MgCl$_2$, or MnCl$_2$ are used\cite{Parsegian92,Gelbart03}. 
This shows that the hydration
effect and the entropy of the hydrated water molecules are
significant and need to be properly taken into account when
one deals with the problem of DNA confinement inside viral capsids. 
In this section, a first step is taken to study this ion specific
effect. Specifically, we study how DNA$-$DNA interaction is affected
by changing the radius of the counterions. 

\begin{figure}[ht]
\resizebox{8cm}{!}{\includegraphics{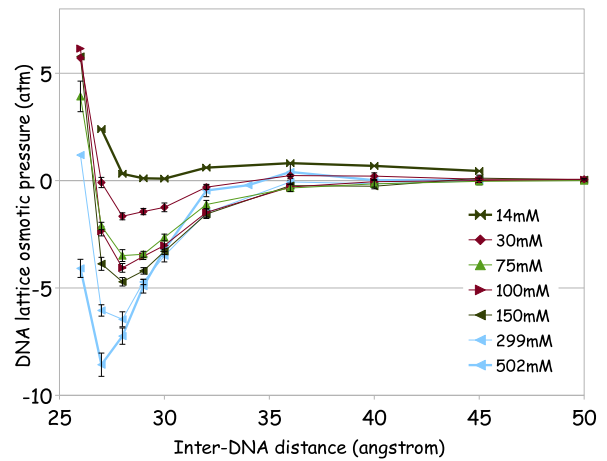}}
\caption{(Color online) The osmotic pressure of the DNA bundle
as function of the interaxial DNA distance $d$ for different
divalent counterion concentration $c_Z$ shown in the inset. 
The solid lines are guides to the eye. The counterion radius is
2.5 \AA
}
\label{fig:posm25}
\end{figure}

\begin{figure}[ht]
\resizebox{8cm}{!}{\includegraphics{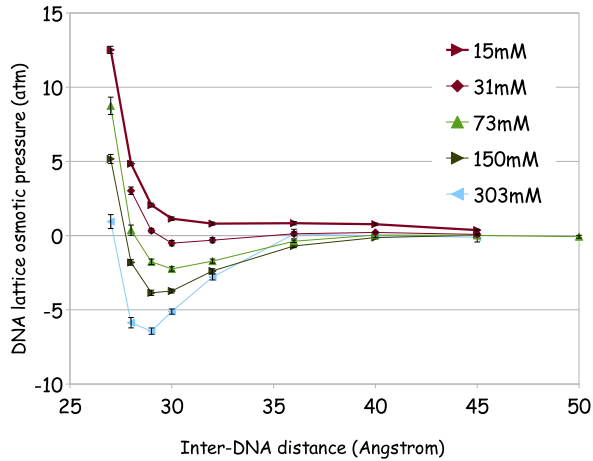}}
\caption{(Color online) The osmotic pressure of the DNA bundle
as function of the interaxial DNA distance $d$ for different
divalent counterion concentration $c_Z$ shown in the inset. 
The solid lines are guides to the eye. The counterion radius is
$\sigma_Z = 3.0$\AA
}
\label{fig:posm30}
\end{figure}

In Fig. \ref{fig:posm25}, and Fig. \ref{fig:posm30}, the dependence of DNA-DNA
"effective" interaction on the DNA-DNA separation distance are plotted
for the counterion radii 2.5\AA\ and 3\AA\ respectively. Compare to 
similar plot for the case of $\sigma_Z = 2.0$\AA\ (Fig. \ref{fig:posm20}),
we can clearly see that the main physics remains when we change
the counterion size. The DNA-DNA short-range interaction remains
evident. However, the depth and location of the strongest attraction
change when the counterion size changes. The smallest counterions
(2\AA) cause the strongest attraction among DNA at smaller distance.
This is easily understood, the smaller counterion cause less entropic
cost of bringing DNA closer to each other. Hence the short-range
attraction is enhanced.

\begin{figure}[ht]

%	\centering
%	\begin{subfigure}{.5\textwidth}
%		\centering
%		\includegraphics[width=.4\linewidth]{PvD_Rdependence.eps}
\resizebox*{8cm}{!}{\includegraphics{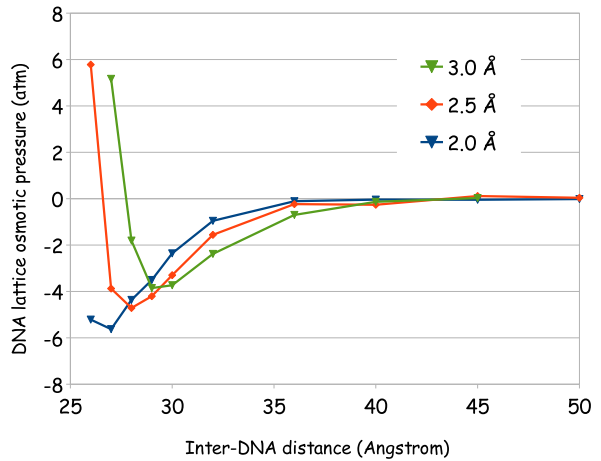}}
%		\caption{b)}
%		\label{fig:sub1}
%	\end{subfigure}%
%	\begin{subfigure}{.5\textwidth}
%		\centering
%		\includegraphics[width=.4\linewidth]{PvD_Rdependence_shifted.eps}
\resizebox*{8cm}{!}{\includegraphics{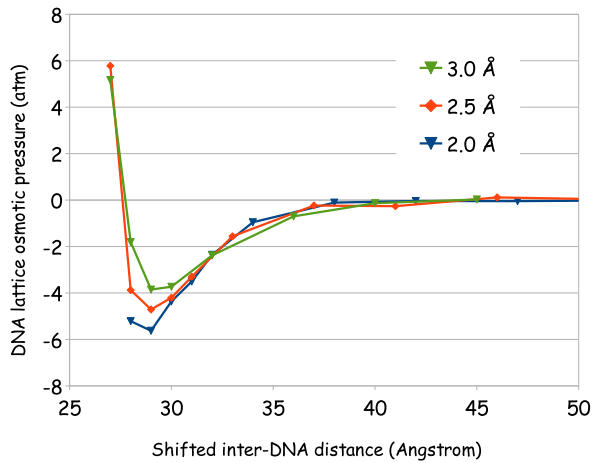}}
%		\caption{a)}
%		\label{fig:sub2}
%	\end{subfigure}
%
\justifying	
\caption{(Color online) a) The osmotic pressure of the DNA hexagonal bundle as function of the lattice constant, $d$, for three values
of the counterion radius, at the same counterion concentration
of 150mM. b) The same plot with the horizontal axis shifted
by $2\sigma_Z$ showing a good degree of overlapping of the three curves
with regard to the equilibirum position and the attractive electrostatic interaction. }
\label{fig:equipos}
\end{figure}

The change in the equilibrium separation of DNA in the bundle is
even more evident in Fig. \ref{fig:equipos}a. In this figure, the osmotic pressure 
(which is proportional to the effective DNA$-$DNA interaction)
of the hexagonal DNA bundle is plotted
as a function of the inter DNA distance for three counterion
sizes, 2\AA, 2.5\AA, and 3\AA, respectively. The counterion concentration
is chosen to be approximately 150mM in each simulation.
As one can see, the first consequence of changing counterion size is obviously
the equilibrium distance of the DNA bundle. The optimal inter DNA
distance, $d^*$, where the short range DNA attraction is strongest
increases with the counterion radius. As the counterion radius is increased
from 2.0\AA\ to 2.5\AA\ to 3.0\AA\, $d^*$ increases from 26\AA\ to 27\AA\ then
29\AA\ respectively.

However, it is an interesting observation that not only the optimum distance
$d^*$ is shifted by $2\sigma_Z$, the interaction between 
DNA molecules from the distance
$d^*$ to $\infty$, which is dominated by electrostatics, is shifted by 
the same amount. This is evident as in Fig. \ref{fig:equipos}b where
the horizontal axis for each curve is shifted by $2\sigma_Z$.
One can see that the right side of these curve from the
distance $d^*$ to $\infty$ show a good degree of overlapping.

This is in agreement with the
``correlated liquid" nature of DNA$-$DNA attraction mediated by
multivalent counterions \cite{NguyenRMP2002,NguyenJBP2013}. In this
strongly correlated liquid theory of DNA$-$DNA interaction, 
the combined system of DNA$+$condensed counterions acts as a charged 
metallic cylinder. The correlations between the condensed counterions on the
surface of two neighboring DNA induce a short range attraction
between them. In this theory, the center of mass
of condensed counterion cannot approach the DNA surface at a distance
less than its radius, $\sigma_Z$. Because of this, the {\em effective} 
surface of the dressed metallic DNA is lifted off the {\em bare} DNA surface by a distance of 
\beq
x = \sigma_Z+\lambda+|\xi|,
\eeq
where $\lambda$ is the Goy-Chapman length.
The length $\xi$ is half the (negative) screening length of the strongly
correlated liquid of the condensed counterions on the surface of the
DNA molecule.
\beq
\xi = \frac{\varepsilon}{4\pi (Ze)^2} \frac{d\mu}{dn}
\eeq
with $\mu$ the chemical potential of a counterion in the liquid,
and $n$ is its two-dimensional density. This screening length, $|\xi|$, 
depends weakly on the ratio, $\sigma_Z/r_{DNA}$. For our purpose, 
it can be considered to be constant.
Therefore, if one considers the correlation-induced attraction between
two DNA cylinders only works when the closest approach between
their surfaces is greater than $2x$ (so that the two DNA's "effective" 
metallic layers donot overlapped), one immediately comes to
the conclusion that the electrostatic like-charged attraction between 
two neighboring DNA cylinders is simply shifted by a distance of 
$2\sigma_Z$ when the radius of the counterion changes. 
This agrees with our simulation results.

\begin{figure}[ht]
\resizebox*{8cm}{!}{\includegraphics{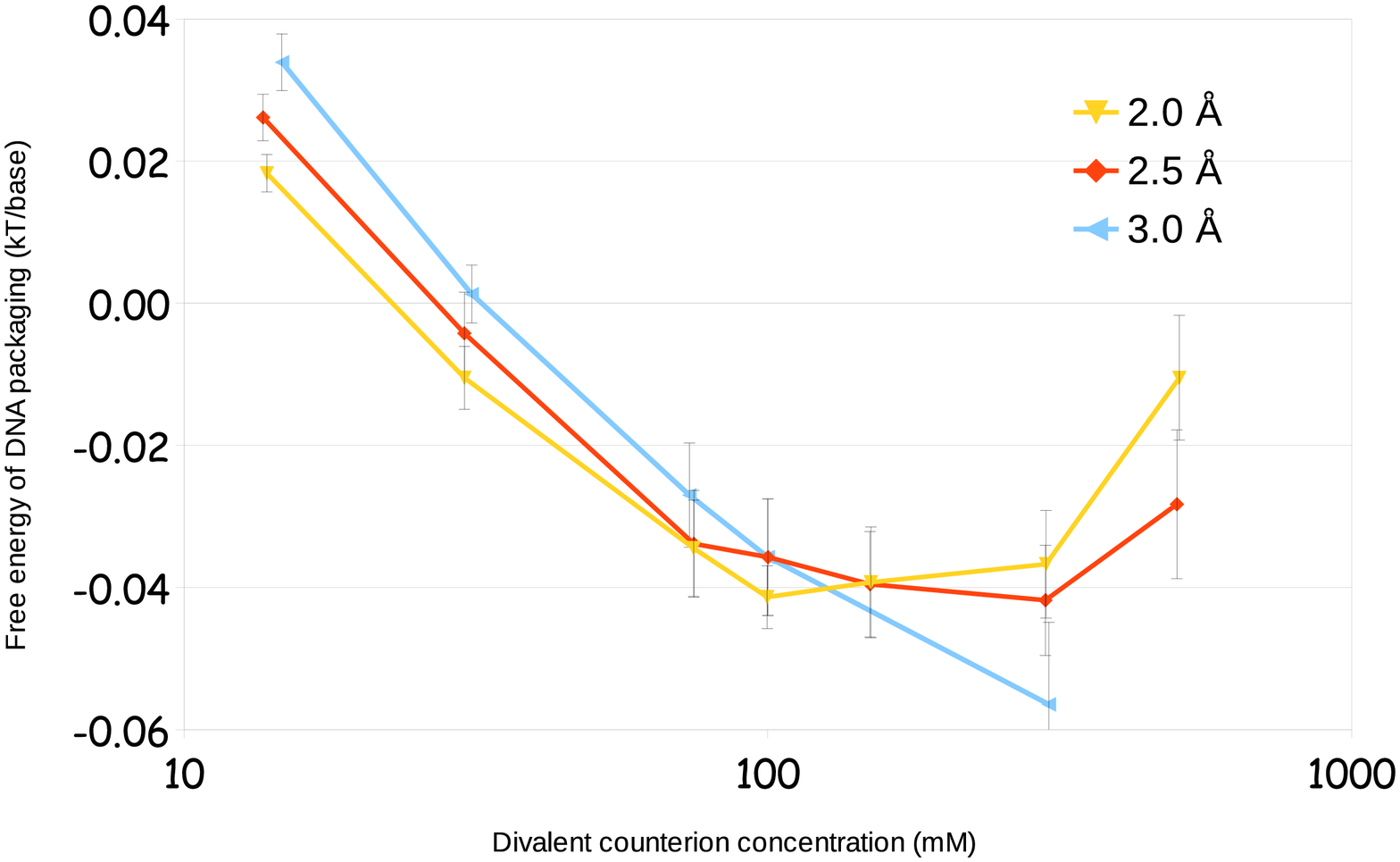}}	
\caption{Free energy per nucleotide base of packaging DNA molecules into
bundles as a function of counterion concentration $c_Z$ for different 
radii of the counterions.}
\label{fig:energypac}
\end{figure}

In Fig. \ref{fig:energypac}, the free energy of packaging DNA into
an hexagonal bundle with the optimal inter$-$DNA distance, $d^*$,
is plotted as a function of the counterion concentrations for the
three different counterion radii.
It can be seen clearly that, within the range of counterion concentration studied,
there is a quantitative and qualitative 
difference in the free energy of packaging for the three sizes
of counterion consider. For $\sigma_Z = 2$\AA\ and 2.5\AA,
the dependence of the free 
energy of packaging DNA in bundle on the concentration $c_Z$ is non-monotonic.
However, for the larger counterion size, $\sigma_Z = 3$\AA,
in the range of concentration considered, DNA condense later but stronger
into hexagonal bundle as the counterion concentration increases.
This behavior is actually observed in experiments. While the
non-monotonic behaviors of DNA ejection is observed clearly for
MgSO$_4$ salt, and somewhat evident for MgCl$_2$ salt,
MnCl$_2$ are known to condense DNA in free solution without ever disintegrated
\cite{Parsegian92,Knobler08}.
Our simulation suggests that the difference in the hydration radius of the counterions
can be used to explain such differences. Our results suggests that Mn$^{2+}$ counterion has
larger ion radius.  This is in good qualitative agreement with computational and EXAFS and X$-$ray studies
on divalent counterions hydration shell (see Table 3 of reference \onlinecite{RCl2hydrationMD} and
the corresponding references therein). These work shown 
that the number of water molecules in the hydration
shell of ions increases with its atomic number. Specifically, as the atomic
number of the divalent counterions increases from Mg$^{2+}$, 
Ca$^{2+}$, Sr$^{2+}$ to Ba$^{2+}$, the coordination number
increases from 6 to 9 water molecules in the hydration shell. Even though,
Mn$^{2+}$ hydration was not studied in these works, its atomic
number is higher than that of Ca$^{2+}$ and Mg$^{2+}$ ions suggesting 
that its hydration radius is larger than that of Mg$^{2+}$ counterions.

It is of importance to note that, according to our Fig. \ref{fig:energypac}, 
although the larger counterions do not produce
a reentrant non-monotonic behavior, they actually cause
stronger DNA-DNA attraction energy. Based on what is observed
from Fig. \ref{fig:posm20}, Fig. \ref{fig:posm25}, Fig. \ref{fig:posm30}
and the horizontally shifted Fig. \ref{fig:equipos}b, this observation can be
explained as the result of two effects. First, smaller counterions
can condense better on DNA, causing a stronger short-range like charge attraction 
among DNA cylinders. However they also cause a higher degree of 
overcharging at larger concentrations, so it is costlier to packaging
DNA. This is evident by the increase in 
the packaging free energy at higher concentration for $\sigma_Z =2$\AA. 
Secondly, the short-range attraction between DNA is shifted to 
larger $d$ for larger counterions. Since one integrates $\int PdV$ 
to find the packaging free energy, a simple geometric argument shows that
the contribution from larger $d$ would dominate this integral, 
therefore the larger counterions can cause lower energy minimum at large concentration.

\section{Conclusion\label{sec:conclusion}}
In this paper, we use a Grand-Canonical Monte-Carlo simulation to study 
the electrostatics of DNA condensation, using a primitive model
for the screen ions. Specifically, the effective electrostatic 
interaction between DNA molecules in a hexagonal bundle is computed 
in the presence of  50mM monovalent counterions and with varying concentration of 
divalent counterions. The entropy of DNA configure fluctuation
is suppressed in simulation by fixing the position of the DNA cylinders
in the bundle. Such study can be applied directly
to the experimental problem of DNA ejection from bacteriophages where
DNA condensed in a strongly confined environment. 
It is shown that, even at the level of non-specific electrostatic
interaction, divalent counterions can strongly influence DNA interaction
and packaging.
The simulation results for divalent counterions with 2.0\AA\ radius show that
the electrostatic free energy of packaging DNA into hexagonal bundle
varies non-monotonically with the counterion concentration. However,
divalent counterions donot correlate strong enough with each other
to drive DNA de-condensation.
% Since DNA cylinders are rigid in our simulation,
% our system is appropriate to the packaging of DNA in a confined
% space like inside viral capsid.
% The non-monotonic behavior agrees well with
% the strongly correlated liquid theory of DNA$-$DNA interaction in the
% presence of multivalent counterions and with experiment of ejection
% of DNA from bacteriophages. 

The counterion specificity such as the ion hydration radius
can influence strongly the qualitative and quantitative
picture of DNA condensation. Three different counterion sizes
are studied. They show that the non-monotonicity changes significantly
and disappears as the counterion size increases. 
The most important results of this paper are presented in 
Fig. \ref{fig:equipos} and Fig. \ref{fig:energypac}, where
it is shown that increasing counterion radius simply raises 
the "metallic" surface of condensed counterions off the DNA and shift the 
correlation-induced attraction between two DNA cylinders by
an amount of $2\sigma_Z$. This interestingly is responsible for making
the larger counterions to cause a deeper minimum of DNA packaging
free energy. In fact, in the range of concentration considered in our simulation with counterion
radius of 3\AA, this free energy keeps going lower with increasing
 counterion concentration. Such qualitative differences are 
observed with DNA condensation experiments involving Mg$^{2+}$ and
Mn$^{2+}$ counterions and suggesting that Mn$2+$ has bigger ion radius,
in agreement with previous computational and EXAFS and X$-$ray experimental results.
 
Going beyond the scope of DNA ejection experiments,
we believe the quantitative results of our paper can be used 
to understand many other experiments involving DNA and divalent counterions.

\begin{acknowledgments}
We would like to thank Lyubartsev, Shklovskii, Evilevich,
Fang, Gelbart for valuable discussions. 
TTN acknowledges the financial support of
the Vietnam National Foundation for Science and Technology
NAFOSTED Contract 103.02-2012.75 and the USA National Science Foundation 
grant NSF CBET-1134398. The authors are indebted to 
A. Lyubartsev for providing us with the 
source code of their Expanded Ensemble Method.
\end{acknowledgments}

\bibliography{nttpaper}

\end{document}